\documentclass{article}
\usepackage{spconf} 
\usepackage{amsmath,amssymb}
\usepackage{graphicx}
\usepackage{bm,cite,mathtools}
\usepackage{subfig}
\usepackage{siunitx}
\newcommand{\mR}{\mathbb{R}}
\newcommand{\mC}{\mathbb{C}}
\newcommand{\mE}{\mathbb{E}}
\newcommand{\e}{\bm{e}}
\renewcommand{\d}{\bm{d}}
\renewcommand{\r}{\bm{r}}
\newcommand{\x}{\bm{x}}
\newcommand{\y}{\bm{y}}
\newcommand{\G}{\bm{G}}
\newcommand{\W}{\bm{W}}
\newcommand{\A}{\bm{A}}

\renewcommand{\H}{\mathsf{H}}
\newcommand{\Trans}{\mathsf{T}}
\newcommand{\ue}{u_{\mathrm{e}}}
\newcommand{\up}{u_{\mathrm{p}}}
\newcommand{\us}{u_{\mathrm{s}}}

\newcommand{\bmzeta}{\bm{\zeta}_{y}}
\newcommand{\bmzx}{\bm{z}_{x}}
\newcommand{\Wf}{\bm{W}_{\mathrm{fixed}}}
\newcommand{\Je}{J_{\mathrm{e}}}
\newcommand{\Jpe}{J_{\mathrm{PE}}}
\newcommand{\Jtrans}{J_{\mathrm{trans}, n}}
\newcommand{\Pred}{P_{\mathrm{red}}}
\title{Spatial active noise control method based on sound field interpolation from reference microphone signals}
\name{Kazuyuki Arikawa, Shoichi Koyama, and Hiroshi Saruwatari}
\address{The University of Tokyo, 7-3-1 Hongo, Bunkyo-ku, Tokyo 113-8656, Japan}
\begin{document}
\ninept
\maketitle
\begin{abstract}
A spatial active noise control (ANC) method based on the interpolation of a sound field from reference microphone signals is proposed. In most current spatial ANC methods, a sufficient number of error microphones are required to reduce noise over the target region because the sound field is estimated from error microphone signals. However, in practical applications, it is preferable that the number of error microphones is as small as possible to keep a space in the target region for ANC users. We propose to interpolate the sound field using reference microphones, which are normally placed outside the target region, instead of the error microphones. We derive a fixed filter for spatial noise reduction on the basis of the kernel ridge regression for sound field interpolation. 
Furthermore, to compensate for estimation errors, we combine the proposed fixed filter with multichannel ANC based on a transition of the control filter using the error microphone signals.
Numerical experimental results indicate that regional noise can be sufficiently reduced by the proposed methods even when the number of error microphones is particularly small.
\end{abstract}
\begin{keywords}
spatial active noise control, kernel interpolation, adaptive filtering algorithm, sound field control
\end{keywords}
\section{Introduction} 
\label{sec:intro}
\vspace{-2.5mm}
Active noise control (ANC) is used to reduce unwanted primary noise by driving secondary sources (loudspeakers). In particular, spatial ANC, which is a technique of suppressing noise in a spatial target region, has attracted attention in recent years, owing to the recent progress in sound field analysis and synthesis methods~\cite{poletti2005three,spors2008theory,wu2008theory,Koyama:IEEE_J_ASLP2013,Ueno:IEEE_SPL2018,Ueno:IEEE_J_SP_2021}. 
Although conventional multichannel ANC methods for minimizing the power of error microphone signals~\cite{Elliott:LMS, kuo1999active} can be applied to regional noise reduction, there is no guarantee that the primary noise is reduced in the region between the error microphone positions. 
To achieve noise reduction over a spatial target region, 
several spatial ANC methods have been proposed on the basis of the spherical/circular harmonic expansion~\cite{Zhang:ANC2018, Bu:ACM2018, maeno2019spherical, Sun:ICASSP2019} or kernel interpolation of the sound field~\cite{Ito:ICASSP2019,Koyama:IEEE_ACM_J_ASLP2021}. 
In these methods, adaptive filtering algorithms to minimize regional noise power are obtained on the basis of the sound field estimated using error microphones. 
To fully capture the sound field inside the target region, it is required to place a sufficient number of error microphones inside or near the target region. 

Considering practical spatial ANC applications, e.g., inside automobile cabins~\cite{samarasinghe2016recent} and office rooms~\cite{lam2020active}, ANC users normally exist in the target region. Therefore, it will be difficult to place many microphones inside or near the target region and it is preferable that the number of error microphones is as small as possible to keep a space for the users. 
Since most current spatial ANC methods are based on the sound field estimated from error microphone signals, they are not applicable to suppressing spatial noise sufficiently 
 over the target region when the number of error microphones is particularly small.

We propose a spatial ANC method based on the sound field estimated using reference microphones, instead of error microphones. In typical feedforward ANC systems, reference microphones are placed outside the target region to capture the incoming primary noise signals. By using a relatively large number of reference microphones, we can assume that the sound field produced by primary noise sources, i.e., the primary noise field, can be estimated from reference microphone signals. 
The error microphones are not used to estimate the spatial sound field; thus a space for ANC users can be saved by reducing the number of error microphones.
A side benefit of this setting is that effects from scattering waves produced by obstacles in the target region can be alleviated because they will be negligible at the reference microphone positions. Such an ANC system will be easy to use because the system scale in the target region is small. 

We formulate the spatial ANC method with the above-described concept based on the kernel ridge regression for sound field interpolation~\cite{Ueno:IWAENC2018,Ueno:IEEE_J_SP_2021}.
A major advantage of the kernel ridge regression against the spherical/cylindrical harmonic analysis is its applicability to arbitrary microphone array geometry.
Based on the estimated sound field, we derive a fixed filter for minimizing the acoustic potential energy in the target region. In addition, to compensate for the estimation error using the error microphone signals, a normalized least-mean-square (NLMS) algorithm for transitioning from the fixed filter to the adaptive control filter for multichannel ANC is derived. We evaluate the spatial noise reduction performance of the proposed methods by numerical experiments in a two-dimensional (2D) space including a scattering object. For simplicity, all the formulations are derived in the frequency domain; however, time-domain algorithms can be readily derived in a similar manner to our previous work~\cite{Koyama:IEEE_ACM_J_ASLP2021}. 
\vspace{-2.5mm}
\section{Problem statement and prior works}
\label{sec:preli}
\vspace{-2.5mm}
\subsection{Problem statement}

\begin{figure}[tb]
  \centering
  \centerline{\includegraphics[width=0.82\columnwidth]{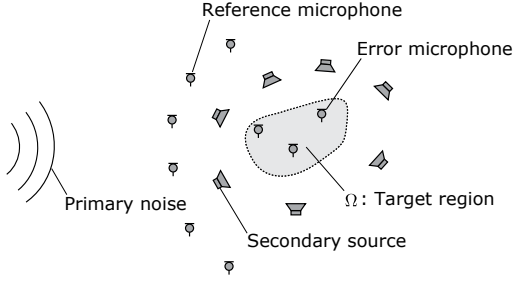}}
\caption{Schematic diagram of feedforward spatial ANC with small number of error microphones.}
\label{fig:1-01}
\end{figure}

The objective of spatial ANC is to reduce the incoming primary noise over $\Omega \subset \mR^D$, where $D=2$ or $3$, by driving multiple secondary sources. As shown in Fig.~\ref{fig:1-01}, error microphones are placed inside $\Omega$ to capture the sound field, and secondary sources and reference microphones are placed outside $\Omega$. In particular, we here assume that the number of error microphones is small whereas the number of reference microphones is relatively large. 

The numbers of error microphones, secondary sources, and reference microphones are denoted as $M$, $L$, and $R$, respectively. 
We denote the driving signals of the secondary sources and the observed signals of the reference and error microphones at the time frame $n$ and angular frequency $\omega$ as $\y_n(\omega) \in \mC^L$, $\x_n(\omega) \in \mC^R$, and $\e_n(\omega) \in \mC^M$, respectively. Hereafter, the argument $\omega$ is omitted for notational simplicity. By denoting the primary noise source signals at the error microphone positions as $\d_n(\omega) \in \mC^M$, we can represent the error microphone signals as
\begin{align}
    \e_n = \d_n + \G\y_n, \label{eq:1-01}
\end{align}
where $\G \in \mC^{M \times L}$ consists of transfer functions from the secondary sources to the error microphones. We assume that $\G$ is given by measuring or modeling them in advance. 
The driving signals of the secondary sources are obtained by filtering the reference microphone signals with the control filter matrix $\W_n \in \mC^{L \times R}$ as
\begin{align}
    \y_n = \W_n \x_n.
\end{align}
For simplicity, we assume that the secondary source signals at the reference microphone positions, i.e., \textit{acoustic feedback}, are negligible. 
The effect of acoustic feedback can be mitigated by using directional loudspeakers~\cite{murao2012basic} or by cancelling acoustic feedback using the estimated acoustic feedback path~\cite{akhtar2007active}. 

\subsection{Prior works on spatial ANC}
\label{subsec:preli-prior}

When applying a conventional multichannel ANC method to the spatial ANC, the cost function is defined as the expected value of the power of the error microphone signals as
\begin{align}
    \Je = \mE\left[\|\e_n\|_2^2\right]. \label{eq:2.2-01}
\end{align}
By replacing the expected value with the instantaneous value at the time frame $n$, we can derive the NLMS algorithm for updating $\W_n$ to minimize $\Je \simeq \|\e_n\|_2^2$ as
\begin{align}
    \W_{n+1} &= \W_n - \mu_n \frac{\partial}{\partial \W_n^{\ast}}\|\e_n\|_2^2 \notag \\
    &= \W_n - \mu_n \G^{\H}\e_n\x_n^{\H},
\end{align}
where $(\cdot)^{\ast}$ and $(\cdot)^{\H}$ denote the complex conjugate and conjugate transpose, respectively. 
Here, $\mu_n$ is the step size obtained as
\begin{align}
\mu_n = \frac{\mu_0}{\|\G^{\H}\G\|_2\|\x_n\|_2^2 + \epsilon}, \label{eq:2.2-03}
\end{align}
where $\|\cdot\|_2$ for matrices denotes the maximum singular value, $\mu_0 \in (0, 2)$ is a normalized step-size parameter, and $\epsilon > 0$ is a regularization parameter to avoid zero division. 
The initial control filter $\W_{0}$ is typically set to $\bm{0}$.

Since the cost function defined in \eqref{eq:2.2-01} evaluates the power of error microphone signals, the primary noise can be suppressed only around the error microphone positions. In spatial ANC methods, the cost function is typically defined as the acoustic potential energy inside $\Omega$, represented as
\begin{align}
    \Jpe = \mE\left[\int_{\Omega}|\ue(\r, n)|^2\, \mathrm{d}\r\right] \label{eq:2.2-04},
\end{align}
where $\ue(\cdot, n)$ is the total sound field generated by the primary noise and secondary sources at the time frame $n$. 
In the spatial ANC methods, $\ue(\r, n)$ $(\r\in\Omega)$ and $\Jpe$ are typically estimated from $\e_n$. However, when the number of error microphones is particularly small, it is significantly difficult to estimate the sound field over the target region. Therefore, the current spatial methods are not suitable for this setting. 
\vspace{-2.5mm}
\section{Spatial ANC method based on kernel interpolation from reference microphone signals}
\vspace{-2.5mm}
As discussed in Sect.~\ref{subsec:preli-prior}, the current spatial ANC methods based on kernel interpolation are not suitable when the number of error microphones is particularly small because the sound field is interpolated from the error microphone signals. The conventional multichannel ANC method can be applied, but there is no guarantee that the primary noise is suppressed over the target region. We propose a spatial ANC method on the basis of the primary noise field estimated using reference microphones, instead of error microphones, by assuming that the number of reference microphones is relatively large. 
By reducing the number of error microphones, we can save a space for the ANC users.
We formulate the acoustic potential energy on the basis of the kernel interpolation of the primary noise field from the reference microphone signals. Then, two algorithms are derived: a fixed filter for minimizing the acoustic potential energy and an NLMS algorithm for transitioning from the fixed filter to the control filter for multichannel ANC. 

\subsection{Kernel interpolation of primary noise field from reference microphones}\label{subsec:3-1}

First, we consider the interpolation of the primary noise field inside $\Omega$ using kernel ridge regression. We here assume that an approximate direction of a single primary noise source is given and denoted as $\bm{\eta}$. 
The primary noise field at the time frame $n$, denoted as $\up(\cdot, n)$, is estimated from the reference microphone signals as
\begin{align}
    \up(\r, n) = \bmzx(\r)^{\Trans}\x_n \label{eq:3.1-01}
\end{align}
 with the interpolation filter
 \begin{align}
     \bmzx(\r) = \left[(\bm{K} + \lambda \bm{I}_R)^{-1}\right]^{\Trans}\bm{\kappa}(\r).
 \end{align}
 Here, $\bm{K} \in \mC^{R \times R}$ and $\bm{\kappa}(\r) \in \mC^{R}$ are the Gram matrix and the vector consisting of the kernel function $\kappa(\cdot, \cdot)$, respectively. 
 We apply the kernel function with directional weighting~\cite{Ueno:IEEE_J_SP_2021, Ito:ICASSP2020}, which is defined as
 \begin{align}
     \kappa(\r_1, \r_2) = \begin{cases}
         J_0 \left( \sqrt{(\mathrm{j}\beta \bm{\eta} - k\r_{12})^{\Trans}(\mathrm{j}\beta \bm{\eta} - k\r_{12})} \right) & \text{if $D = 2$} \\ 
         j_0 \left( \sqrt{(\mathrm{j}\beta \bm{\eta} - k\r_{12})^{\Trans}(\mathrm{j}\beta \bm{\eta} - k\r_{12})} \right) & \text{if $D = 3$}
     \end{cases}, \label{eq:3.1-03}
 \end{align}
 where $\r_{12} = \r_1 - \r_2$, $k$ is the wave number, and $J_0(\cdot)$ and $j_0(\cdot)$ are the $0$th-order Bessel and spherical Bessel functions of the first kind, respectively. The scalar $\beta \geq 0$ is a weighting parameter for controlling the sharpness of weighting.
 Even when the primary noise source direction is unknown, 
 we can also apply the kernel function in \eqref{eq:3.1-03}
 by setting $\beta$ to $0$. Furthermore, when multiple primary noise sources exist, it is possible to define the kernel function by the weighted sum of \eqref{eq:3.1-03} with different $\beta$ and $\bm{\eta}$ values, and optimize the parameters from the measured signals~\cite{Horiuchi:WASPAA2021}. 
 
\subsection{Fixed filter based on kernel interpolation}
\label{subsec:3-2}

In a manner similar to the primary noise field interpolation described in Sect.~\ref{subsec:3-1}, we estimate the 
sound field produced by the secondary sources, i.e., secondary sound field, as
\begin{align}
    \us(\r, n) = \bmzeta(\r)^{\Trans}\y_n, \label{eq:3.2-01}
\end{align}
where $\us(\cdot, n)$ is the secondary sound field at the time frame $n$ and $\bmzeta(\cdot)$ is the estimation filter.
Although several approaches to derive $\bmzeta(\cdot)$ can be considered, we here assume that $\bmzeta(\cdot)$ is modeled by the free-field Green's function $G(\r, \r^{\prime})$ with the observation position $\r$ and the source position $\r^{\prime}$ as
\begin{align}
    \bmzeta(\r) = 
    \begin{bmatrix}
    G(\r, \r_1) & \cdots & G(\r, \r_L)
    \end{bmatrix}^{\Trans},
\end{align}
where $\r_l$ is the $l$th secondary source position. 
With this assumption, we do not need to measure the secondary path. 
Furthermore, when the transfer functions from secondary sources to multiple points in $\Omega$ are given, it is also possible to construct the interpolation filter $\bmzeta(\cdot)$ based on the kernel ridge regression~\cite{Arikawa:ICASSP2022}, which can improve the interpolation accuracy in a reverberant environment.


By adding the primary noise and secondary sound fields respectively estimated using \eqref{eq:3.1-01} and \eqref{eq:3.2-01}, we can estimate the total sound field inside $\Omega$, $\ue(\cdot, n)$, as
\begin{align}
    \ue(\r, n) &= \up(\r, n) + \us(\r, n) \notag \\
    &= \bmzx(\r)^{\Trans}\x_n + \bmzeta(\r)^{\Trans}\y_n. \label{eq:3.2-03}
\end{align}
The cost function for spatial ANC is the estimated acoustic potential energy inside $\Omega$, which is defined in \eqref{eq:2.2-04}. 
By substituting \eqref{eq:3.2-03} into \eqref{eq:2.2-04} and by replacing the expected value with the instantaneous value, we can approximate $\Jpe$ as
\begin{align}
    \Jpe &\simeq \int_{\Omega}|\ue(\r, n)|^2\, \mathrm{d}\r \notag \\
    &= \y_n^{\H}\A_{yy}\y_n + \y_n^{\H}\A_{yx}\x_n + \x_n^{\H}\A_{yx}^{\H}\y_n +\x_n^{\H}\A_{xx}\x_n,
\end{align}
where $\A_{yy} \in \mC^{L \times L}$, $\A_{yx} \in \mC^{L \times R}$, and $\A_{xx} \in \mC^{R \times R}$  are interpolation matrices defined as
\begin{align}
    \A_{yy} &= \int_{\Omega}\bmzeta(\r)^{\ast}\bmzeta(\r)^{\Trans}\, \mathrm{d}\r \label{eq:3.2-05}\\
    \A_{yx} &= \int_{\Omega}\bmzeta(\r)^{\ast}\bmzx(\r)^{\Trans}\, \mathrm{d}\r \label{eq:3.2-06}\\
    \A_{xx} &= \int_{\Omega}\bmzx(\r)^{\ast}\bmzx(\r)^{\Trans}\, \mathrm{d}\r. \label{eq:3.2-07}
\end{align}
The gradient of $\Jpe$ with respect to the control filter is represented as
\begin{align}
    \frac{\partial \Jpe}{\partial \W_n^{\ast}} \simeq (\A_{yy}\W_n + \A_{yx})\x_n\x_n^{\H}.
\end{align}
When the matrix $\A_{yy}$ is invertible, we can derive the fixed control filter $\Wf$ that satisfies $\partial \Jpe/\partial \W_n^{\ast} = \bm{0}$ for any reference microphone signals as
\begin{align}
    \Wf = -\A_{yy}^{-1}\A_{yx}. \label{eq:3.2-09}
\end{align}
The matrices $\A_{yy}$ and $\A_{yx}$ respectively defined in \eqref{eq:3.2-05} and \eqref{eq:3.2-06} can be calculated offline. Thus, $\Wf$ can be obtained before ANC processing. 
When the matrix $\A_{yy}$ is not invertible, we replace $\A_{yy}^{-1}$ in \eqref{eq:3.2-09} with the pseudo-inverse of $\A_{yy}$. 

\subsection{NLMS algorithm for transitioning from fixed filter}
\label{subsec:3-3}
The fixed filter derived in Sect.~\ref{subsec:3-2} can suffer from estimation errors of the sound field because the number of reference microphones is sometimes insufficient and such microphones are placed far from the target region. Furthermore, the modeling error in the interpolation filter for the secondary sound field, e.g., reverberation, can also cause estimation errors. These errors degrade the noise reduction performance of the fixed filter.

To compensate for the estimation error of the sound field, 
we consider the cost function represented by the time-varying weighted sum of $\Jpe$ and the power of the error microphone signals, $\Je$, as
\begin{align}
    \Jtrans = \gamma^n\Jpe + \Je \label{eq:3.3-01}
\end{align}
with the forgetting factor $\gamma \in (0, 1)$. The initial control filter is set to the fixed filter, i.e., $\W_0=\Wf$. By gradually reducing the weight of $\Jpe$ in the cost function, we expect that the control filter transitions from $\Wf$ to that for multichannel ANC. 
By computing the gradient of $\Jtrans$ with respect to the control filter, we can derive the NLMS algorithm transitioning from the fixed filter as
\begin{align}
    \W_{n+1} &= \W_n - \mu_n \frac{\partial}{\partial \W_n^{\ast}}\Jtrans \notag \\
    &= \W_n - \mu_n\left[\gamma^n(\A_{yy}\y_n + \A_{yx}\x_n) + \G^{\H}\e_n\right]\x^{\H},
\end{align}
where the step size $\mu_n$ is obtained as
\begin{align}
    \mu_n = \frac{\mu_0}{\gamma^n\|\A_{yy}\|_2 + \|\G^{\H}\G\|_2\|\x_n\|_2^2 + \epsilon}. \label{eq:3.3-03}
\end{align}
If $\gamma$ is set to $0$, this algorithm is equivalent to the conventional multichannel ANC method with the initial value $\Wf$. 
\vspace{-2.5mm}
\section{Experiments}
\label{sec:exp}
\vspace{-2.5mm}
\begin{figure}[tb]
  \centering
  \centerline{\includegraphics[width=0.65\columnwidth]{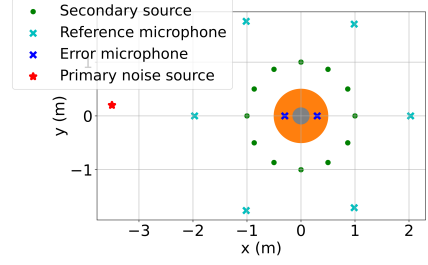}}
\caption{Experimental setup. The orange circular area is the target region, which excludes the area inside the rigid circular object (gray circle). }
\label{fig:4-01}
\end{figure}

We conducted numerical experiments under a 2D free-field condition to evaluate the two proposed methods: a fixed filter based on kernel interpolation from reference microphones (\textbf{Fixed-KIR}) and the NLMS algorithm for transitioning from the fixed filter (\textbf{NLMS w/ Fixed-KIR}). For comparison, we also evaluated the performance of the conventional NLMS algorithm for multichannel ANC (\textbf{NLMS}). 

The target region $\Omega$ was set to a circular region with a radius of \SI{0.5}{m}, whose center was the coordinate origin. $L$ secondary sources and $R$ reference microphones, where $L=12$ and $R=6$, were placed on circles with radii of \SI{1.0}{m} and \SI{2.0}{m}, respectively. To alleviate the forbidden frequency problem~\cite{Koyama:IEEE_ACM_J_ASLP2020}, every reference microphone was shifted \SI{0.03}{m} or $-$\SI{0.03}{m} in 
a radial coordinate. $M=2$ error microphones were placed at (\SI{\pm 0.3}{m}, \SI{0.0}{m}), and a single primary noise source was at (\SI{-3.5}{m}, \SI{0.2}{m}). In addition, as an obstacle inside the target region, a rigid circular object with a radius of \SI{0.15}{m} was placed at the coordinate origin. The overall settings are shown in Fig.~\ref{fig:4-01}.
The primary noise source and all the secondary sources were assumed to be point sources. The noise signal is constant in the frequency domain, and Gaussian noise was added to the reference and error microphone signals at each iteration so that the signal-to-noise ratio became \SI{40}{dB}. 
In \textbf{Fixed-KIR}, the weighing parameter $\beta$ in the kernel function \eqref{eq:3.1-03} used for primary noise field interpolation was set to $6.0$, and the direction $\bm{\eta}$ was set to the actual primary noise source direction. The estimation vector $\bmzeta(\cdot)$ for estimating the secondary sound field was obtained using the 2D free-field Green's function. 
Under this setting, the interpolation matrix $\A_{yy}$ was invertible for all the frequencies used in the experiments, thus, the fixed filter in \eqref{eq:3.2-09} could be calculated.
In \textbf{NLMS w/ Fixed-KIR}, we set the forgetting factor $\gamma$ to $0.9$. The parameters $\mu_0$ and $\epsilon$ in \eqref{eq:2.2-03} and \eqref{eq:3.3-03} were $0.1$ and $10^{-8}$, respectively. 

As a performance measure, we define the regional noise power reduction inside $\Omega$ as 
\begin{align}
    \Pred(n) = 10\log_{10}\frac{\sum_j|\ue(\r_j, n)|^2}{\sum_j|\up(\r_j, n)|^2},
\end{align}
where $\r_j$ is the $j$th evaluation point inside $\Omega$, and $\up(\cdot, n)$ represents the primary noise field at the time frame $n$. We set 556 evaluation points inside $\Omega$, except for the area inside the rigid circular object. 

\begin{figure}[tb]
  \centering
  \centerline{\includegraphics[width=0.82\columnwidth]{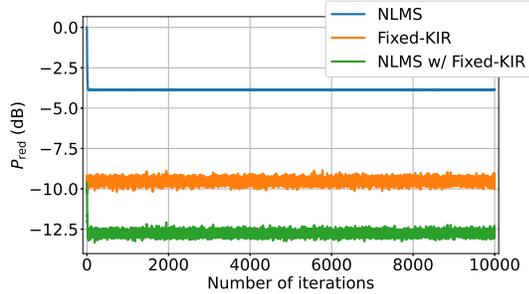}}
\caption{Regional noise power reduction at \SI{400}{Hz}.}
\label{fig:4-02}
\end{figure}

\begin{figure}
    \centering
    \subfloat[][NLMS]{
    \includegraphics[width=0.45\columnwidth]{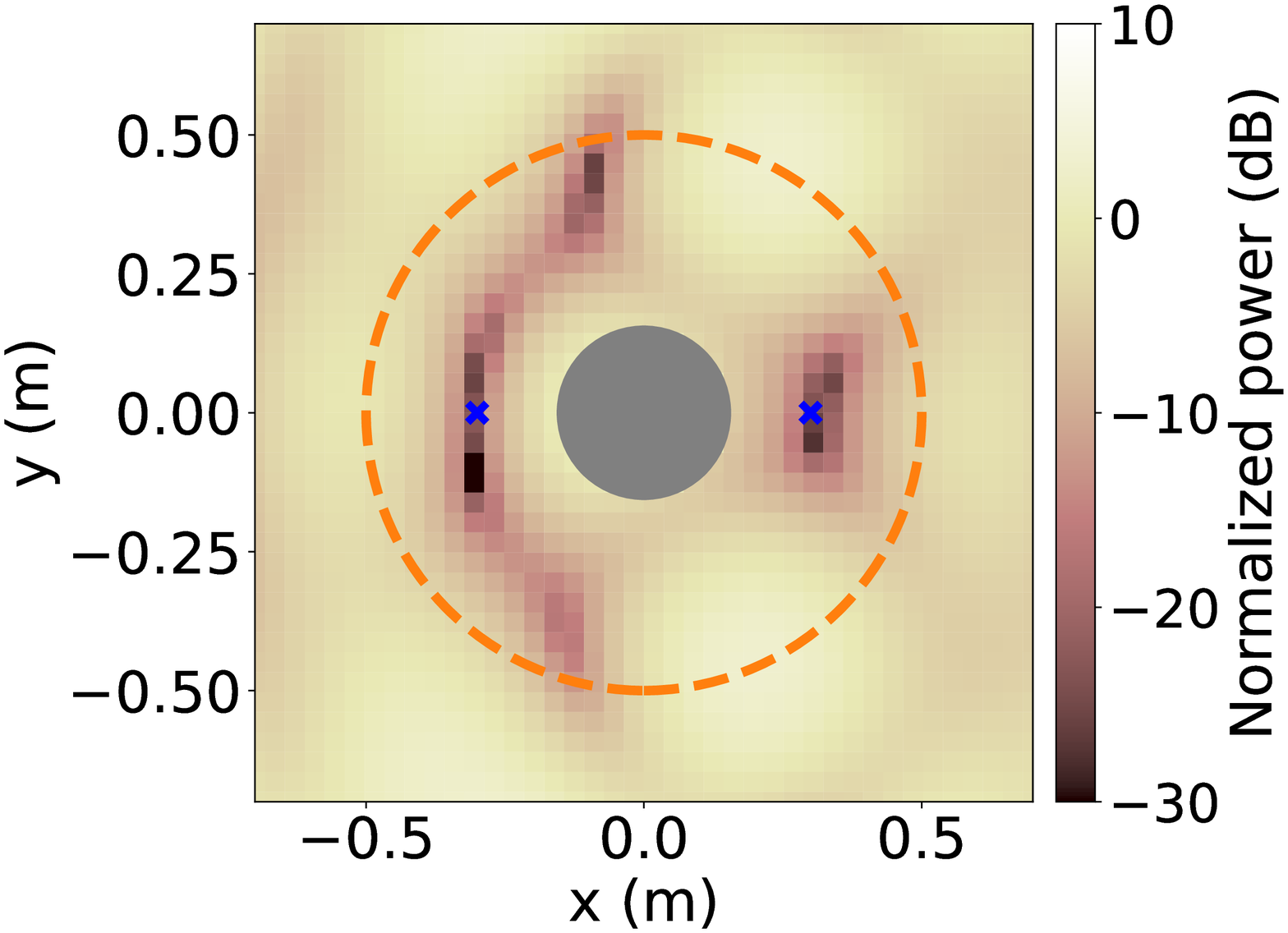}\label{fig:4-03a}} \\
    \subfloat[][Fixd-KIR]{
    \includegraphics[width=0.45\columnwidth]{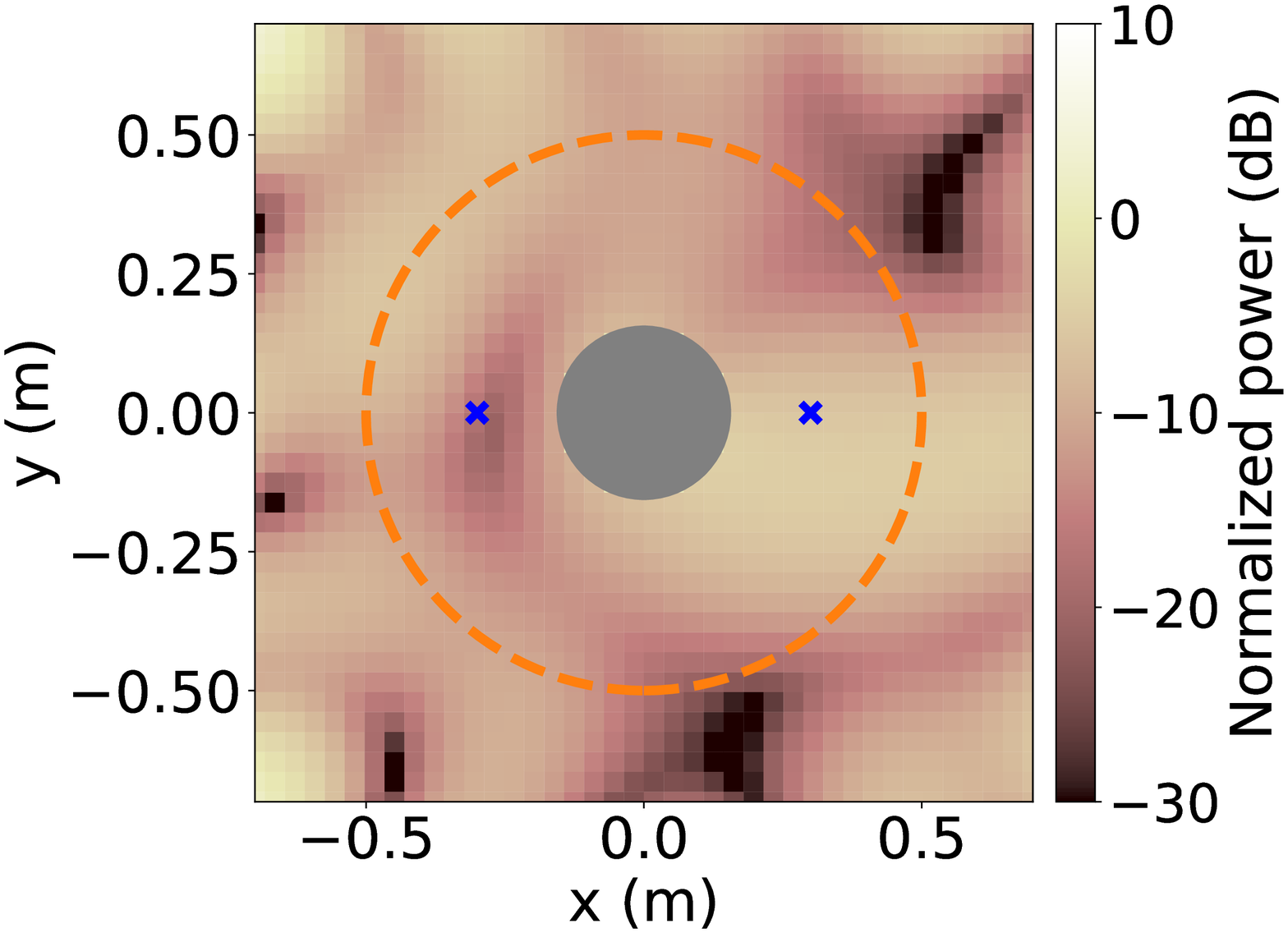}\label{fig:4-03b}}
    \hfill
    \subfloat[][NLMS w/ Fixed-KIR]{
    \includegraphics[width=0.45\columnwidth]{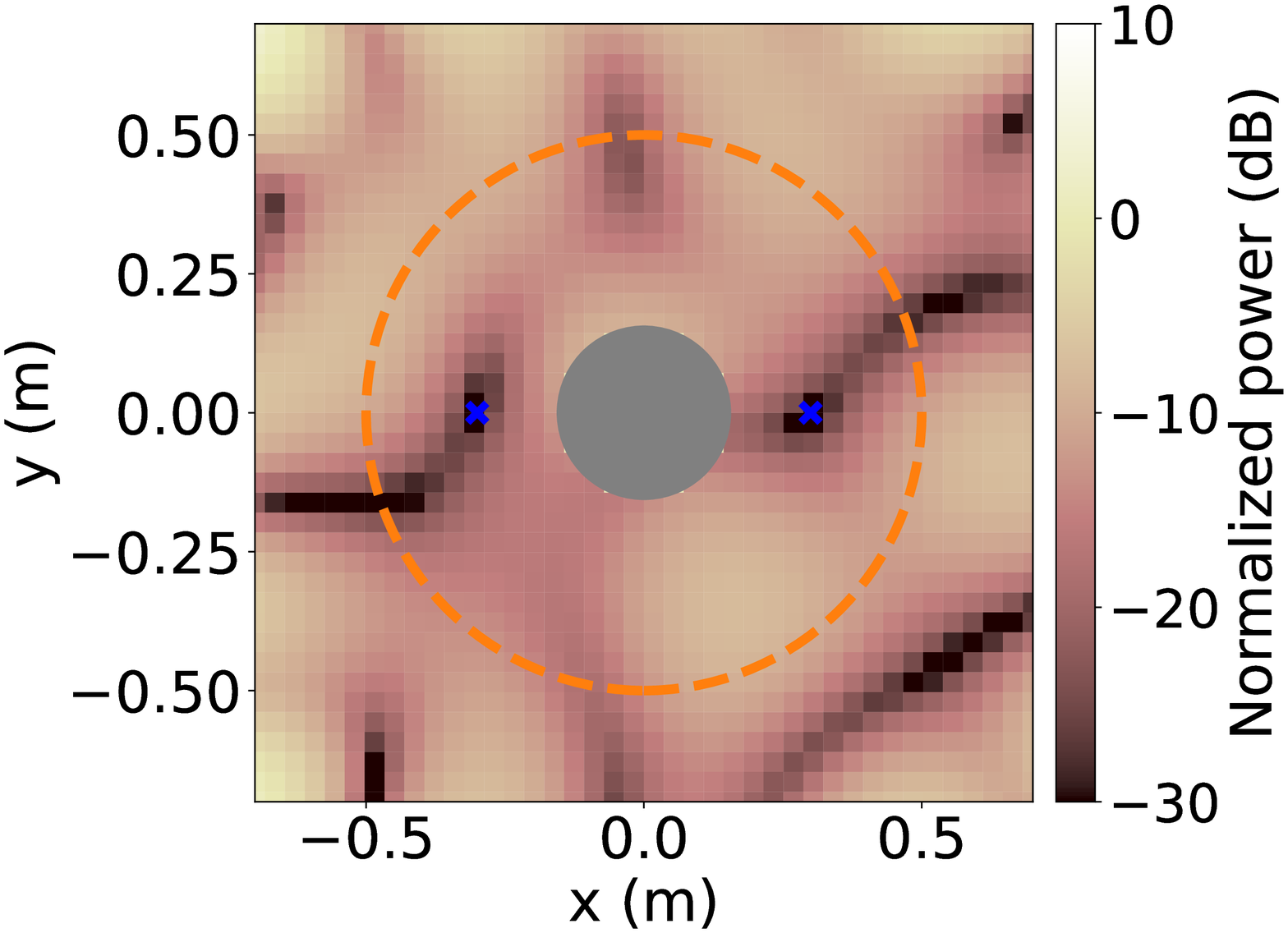}\label{fig:4-03c}}
    \caption{Normalized power distribution at \SI{400}{Hz}. The orange dotted circle indicates the target region, blue crosses are error microphone positions, and the gray circle is the circular object.}
\label{fig:4-03}
\end{figure}

Fig.~\ref{fig:4-02} shows $\Pred$ at each iteration when the noise frequency was \SI{400}{Hz}. 
The two proposed methods achieved a higher noise reduction performance than \textbf{NLMS}. We can also see that the performance of \textbf{NLMS w/ Fixed-KIR} was the same as that of \textbf{Fixed-KIR} at the first iteration and became higher as the adaptation process proceeded.
Fig.~\ref{fig:4-03} shows the power distribution achieved by each method after 10000 iterations, normalized by the average power of the primary noise field inside $\Omega$. 
Fig.~\ref{fig:4-03a} shows that the noise reduction by \textbf{NLMS} was limited to only around the error microphones. In contrast, the noise power over $\Omega$ was reduced by the proposed methods, as can be seen in Figs.~\ref{fig:4-03b} and \ref{fig:4-03c}. 
In particular, \textbf{NLMS w/ Fixed-KIR} achieved a large noise reduction around the error microphone positions while maintaining the noise reduction over $\Omega$. 

\begin{figure}[th]
  \centering
\centerline{\includegraphics[width=0.65\columnwidth]{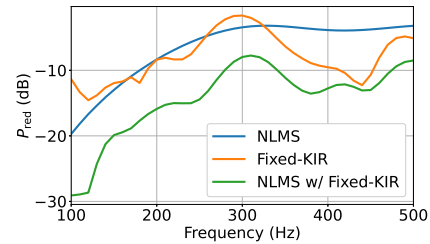}}
\caption{Regional noise power reduction after 10000 iterations with respect to frequency.}
\label{fig:4-04}
\end{figure}
We also varied the noise frequency from \SI{100}{Hz} to \SI{500}{Hz} at \SI{10}{Hz} intervals.
Fig.~\ref{fig:4-04} shows $\Pred$ after 10000 iterations at each frequency. 
The performance of \textbf{Fixed-KIR} was comparable to that of \textbf{NLMS}, but it was worse at some frequencies. 
This is due to the estimation error of the sound field, probably caused by the distant location of the reference microphones from the target region and scattering from the circular object in the target region. 
Moreover, \textbf{NLMS w/ Fixed-KIR} achieved the largest noise reduction at all frequencies, which confirms that transitioning from the fixed filter successfully compensated for the estimation error of the sound field. 
\vspace{-2.5mm}
\section{Conclusion}\label{sec:conclusion}
\vspace{-2.5mm}
We proposed a spatial ANC method based on the kernel interpolation of a sound field from reference microphone signals. Current spatial ANC methods are based on the sound field inside the target region estimated from error microphone signals; however, these methods are not suitable when the number of error microphones is particularly small. In the proposed method, the primary noise field is interpolated using reference microphones instead of error microphones, assuming that the number of reference microphones is relatively large. 
On the basis of the interpolated sound field, a fixed filter minimizing the estimated acoustic potential energy in the target region is derived. 
Furthermore, to compensate for the estimation error of the sound field, we also formulated an NLMS algorithm for transitioning from the fixed filter to a control filter for multichannel ANC. 
The effectiveness of the proposed methods when the number of error microphones is small was validated by numerical experiments. 
\section{Acknowledgment}
\vspace{-2.5mm}
This work was supported by JSPS KAKENHI (Grant Number 22H03608), JST FOREST Program (Grant Number JPMJFR216M, Japan), and Nippon Telegraph and Telephone Corp.

\newpage

\bibliographystyle{IEEE_mod}
\bibliography{str_def_abrv, koyama_en, reference}

\begin{thebibliography}{10}

\bibitem{poletti2005three}
M.~A. Poletti,
\newblock ``Three-dimensional surround sound systems based on spherical
  harmonics,''
\newblock {\em J. Audio Eng. Soc.}, pp. 1004--1025, 2005.

\bibitem{spors2008theory}
S.~Spors, R.~Rabenstein, and J.~Ahrens,
\newblock ``The theory of wave field synthesis revisited,''
\newblock in {\em 124th AES Conv.}, 2008, pp. 17--20.

\bibitem{wu2008theory}
Y.~J. Wu and T.~D. Abhayapala,
\newblock ``Theory and design of soundfield reproduction using continuous
  loudspeaker concept,''
\newblock {\em {IEEE} Trans. Audio, Speech, Lang. Process.}, vol. 17, no. 1,
  pp. 107--116, 2008.

\bibitem{Koyama:IEEE_J_ASLP2013}
S.~Koyama, K.~Furuya, Y.~Hiwasaki, and Y.~Haneda,
\newblock ``Analytical approach to wave field reconstruction filtering in
  spatio-temporal frequency domain,''
\newblock {\em {IEEE} Trans. Audio, Speech, Lang. Process.}, vol. 21, no. 4,
  pp. 685--696, 2013.

\bibitem{Ueno:IEEE_SPL2018}
N.~Ueno, S.~Koyama, and H.~Saruwatari,
\newblock ``Sound field recording using distributed microphones based on
  harmonic analysis of infinite order,''
\newblock {\em {IEEE} Signal Process. Lett.}, vol. 25, no. 1, pp. 135--139,
  2018.

\bibitem{Ueno:IEEE_J_SP_2021}
N.~Ueno, S.~Koyama, and H.~Saruwatari,
\newblock ``Directionally weighted wave field estimation exploiting prior
  information on source direction,''
\newblock {\em {IEEE} Trans. Signal Process.}, vol. 69, pp. 2383--2395, 2021.

\bibitem{Elliott:LMS}
S.~Elliott, I.~Stothers, and P.~Nelson,
\newblock ``A multiple error {LMS} algorithm and its application to the active
  control of sound and vibration,''
\newblock {\em {IEEE} Trans. Acoust., Speech, Signal Process.}, pp. 1423--1434,
  1987.

\bibitem{kuo1999active}
S.~M. Kuo and D.~R. Morgan,
\newblock ``Active noise control: a tutorial review,''
\newblock {\em Proc. IEEE}, pp. 943--973, 1999.

\bibitem{Zhang:ANC2018}
J.~Zhang, T.~D. Abhayapala, W.~Zhang, P.~N. Samarasinghe, and S.~Jiang,
\newblock ``Active noise control over space: A wave domain approach,''
\newblock {\em {IEEE/ACM} Trans. Audio, Speech, Lang. Process.}, pp. 774--786,
  2018.

\bibitem{Bu:ACM2018}
B.~Bu, C.~Bao, and M.~Jia,
\newblock ``Design of a planar first-order loudspeaker array for global active
  noise control,''
\newblock {\em {IEEE/ACM} Trans. Audio, Speech, Lang. Process.}, pp.
  2240--2250, 2018.

\bibitem{maeno2019spherical}
Y.~Maeno, Y.~Mitsufuji, P.~N. Samarasinghe, N.~Murata, and T.~D. Abhayapala,
\newblock ``Spherical-harmonic-domain feedforward active noise control using
  sparse decomposition of reference signals from distributed sensor arrays,''
\newblock {\em {IEEE/ACM} Trans. Audio, Speech, Lang. Process.}, vol. 28, pp.
  656--670, 2019.

\bibitem{Sun:ICASSP2019}
H.~Sun, T.~D. Abhayapala, and P.~N. Samarasinghe,
\newblock ``Time domain spherical harmonic analysis for adaptive noise
  cancellation over a spatial region,''
\newblock in {\em Proc. {IEEE} Int. Conf. Acoust., Speech, Signal Process.
  ({ICASSP})}, 2019, pp. 516--520.

\bibitem{Ito:ICASSP2019}
H.~Ito, S.~Koyama, N.~Ueno, and H.~Saruwatari,
\newblock ``Feedforward spatial active noise control based on kernel
  interpolation of sound field,''
\newblock in {\em Proc. {IEEE} Int. Conf. Acoust., Speech, Signal Process.
  ({ICASSP})}, 2019, pp. 511--515.

\bibitem{Koyama:IEEE_ACM_J_ASLP2021}
S.~Koyama, J.~Brunnstr\"{o}m, H.~Ito, N.~Ueno, and H.~Saruwatari,
\newblock ``Spatial active noise control based on kernel interpolation of sound
  field,''
\newblock {\em {IEEE/ACM} Trans. Audio, Speech, Lang. Process.}, vol. 29, pp.
  3052--3063, 2021.

\bibitem{samarasinghe2016recent}
P.~N. Samarasinghe, W.~Zhang, and T.~D. Abhayapala,
\newblock ``Recent advances in active noise control inside automobile cabins:
  Toward quieter cars,''
\newblock {\em {IEEE} Signal Process. Mag.}, vol. 33, no. 6, pp. 61--73, 2016.

\bibitem{lam2020active}
B.~Lam, D.~Shi, W.-S. Gan, S.~J. Elliott, and M.~Nishimura,
\newblock ``Active control of broadband sound through the open aperture of a
  full-sized domestic window,''
\newblock {\em Sci. rep.}, vol. 10, pp. 1--7, 2020.

\bibitem{Ueno:IWAENC2018}
N.~Ueno, S.~Koyama, and H.~Saruwatari,
\newblock ``Kernel ridge regression with constraint of helmholtz equation for
  sound field interpolation,''
\newblock in {\em Proc. Int. Workshop Acoust. Signal Enhancement ({IWAENC})},
  2018, pp. 436--440.

\bibitem{murao2012basic}
T.~Murao and M.~Nishimura,
\newblock ``Basic study on active acoustic shielding,''
\newblock {\em J. Environ. Eng.}, vol. 7, no. 1, pp. 76--91, 2012.

\bibitem{akhtar2007active}
M.~T. Akhtar, M.~Abe, and M.~Kawamata,
\newblock ``On active noise control systems with online acoustic feedback path
  modeling,''
\newblock {\em {IEEE} Trans. Audio, Speech, Lang. Process.}, vol. 15, no. 2,
  pp. 593--600, 2007.

\bibitem{Ito:ICASSP2020}
H.~Ito, S.~Koyama, N.~Ueno, and H.~Saruwatari,
\newblock ``Spatial active noise control based on kernel interpolation with
  directional weighting,''
\newblock in {\em Proc. {IEEE} Int. Conf. Acoust., Speech, Signal Process.
  ({ICASSP})}, 2020, pp. 8399--8403.

\bibitem{Horiuchi:WASPAA2021}
R.~Horiuchi, S.~Koyama, J.~G.~C. Ribeiro, N.~Ueno, and H.~Saruwatari,
\newblock ``Kernel learning for sound field estimation with l1 and l2
  regularizations,''
\newblock in {\em Proc. {IEEE} Int. Workshop Appl. Signal Process. Audio
  Acoust. ({WASPAA})}, 2021, pp. 261--265.

\bibitem{Arikawa:ICASSP2022}
K.~Arikawa, S.~Koyama, and H.~Saruwatari,
\newblock ``Spatial active noise control based on individual kernel
  interpolation of primary and secondary sound fields,''
\newblock in {\em Proc. {IEEE} Int. Conf. Acoust., Speech, Signal Process.
  ({ICASSP})}, 2022, pp. 1056--1060.

\bibitem{Koyama:IEEE_ACM_J_ASLP2020}
S.~Koyama, G.~Chardon, and L.~Daudet,
\newblock ``Optimizing source and sensor placement for sound field control: An
  overview,''
\newblock {\em {IEEE/ACM} Trans. Audio, Speech, Lang. Process.}, vol. 28, pp.
  686--714, 2020.

\end{thebibliography}
\end{document}